\documentclass[manuscript]{aastex}
\usepackage{graphicx,graphics}
\usepackage{epstopdf}

\shorttitle{Gravitational model of high energy particles}
\shortauthors{de Freitas Pacheco et al.}

\begin{document}


\title{Gravitational model of high energy particles in a collimated jet}


\author{J.A. de Freitas Pacheco$^1$}
\affil{$^1$University of Nice-Sophia Antipolis, Observatoire de la C\^ote d'Azur\\
Laboratoire Lagrange, UMR 7293, BP 4229, 06304 Nice Cedex 4, France}
\email{pacheco@oca.eu}

\author{J. Gariel$^2$ and G. Marcilhacy$^2$}
\affil{$^2$LERMA - UPMC, University Pierre and Marie Curie, Observatoire de Paris\\
CNRS, UMR 8112, 3 rue de Galil\'ee, 94200 Ivry sur Seine, France}
\email{jerome.gariel@upmc.fr,gmarcilhacy@hotmail.com}

\and

\author{N.O. Santos$^{2,3}$}
\affil{$^3$School of Mathematical Sciences, Queen Mary, University of London\\
Mile End Road, London E1 4NS, United Kingdom}
\email{nilton.santos@upmc.fr}


%\altaffiltext{1}{Visiting Astronomer, Cerro Tololo Inter-American Observatory.
%CTIO is operated by AURA, Inc.\ under contract to the National Science
%\altaffiltext{2}{Society of Fellows, Harvard University.}
%\altaffiltext{3}{present address: Center for Astrophysics,
%    60 Garden Street, Cambridge, MA 02138}
%\altaffiltext{4}{Visiting Programmer, Space Telescope Science Institute}
%\altaffiltext{5}{Patron, Alonso's Bar and Grill}


\begin{abstract}

Observations suggest that relativistic particles play a fundamental role in
the dynamics of jets emerging from active galactic nuclei as well as in
their interaction with the intracluster medium. However, no general
consensus exists concerning the acceleration mechanism of those high energy
particles. A gravitational acceleration mechanism is here proposed, in which
particles leaving precise regions within the ergosphere of a rotating
supermassive black hole produce a highly collimated flow. These particles
follow unbound geodesics which are asymptotically parallel to the spin axis of the black
hole and are characterized by the energy $E$, the Carter constant ${\cal Q}$
and zero angular momentum of the component $L_z$. If environmental effects are neglected, the
present model predicts at distances of about 140 kpc from the ergosphere the
presence of electrons with energies around 9.4 GeV. The present mechanism can also
accelerate protons up to the highest energies observed in cosmic rays by the present
experiments.

\end{abstract}


\section{Introduction}

Multiwavelength observations of different astrophysical objects indicate the
presence of ``jets" (Marscher 2005) driven probably by a compact object like
a black hole (BH) or, in some cases, a highly magnetized rotating neutron star.
Jets are observed at scales ranging from sub-parsec up to hundreds of
kiloparsec. Gamma-ray bursts are an example of small scale jets, since they
are supposed to be the consequence of shocks occurring in highly collimated
relativistic flows (M\'esz\'aros et al. 1999; Frail et al. 2001; Rossi et al. 2002)
originated either at the death of a massive star or when a neutron star merges with another neutron
star or with a black hole. Large scale jets are observed in association with
radio-galaxies, quasars, blazars or AGNs (active galactic nuclei) in general and their origin is
probably the consequence of the twisting of magnetic fields anchored in the
very inner region of an accreting disk around a supermassive black hole
(Blandford \& Levinson 1995; Meier et al. 2000; Camenzind 2005).

Jets associated with AGNs have a complex structure with bulk motions characterized
by Lorentz factors typically in the range 10-50 and total power ranging from
$10^{44}$ up to $10^{47}$ ergs$^{-1}$. The composition of these flows is
still uncertain but, in general, supposed to be constituted by electrons and
protons or/and electron-positron pairs or even heavy nuclei.
The composition of the jet is certainly related to the physical processes
that create and energize the flow, representing an important key for the
understanding of the launching mechanism. The large scale acceleration
required to explain the high velocities of the bulk motion cannot be purely
hydrodynamic and are probably a manifestation of the presence of extended
magnetic pressure gradients (Vlahakis \& Konigl 2004). Fully relativistic simulations
of accretion disks around a rotating black hole indicate that unbound flows
can emerge self-consistently from the accretion flow (De Villiers et al. 2005).
According to these simulations, the flow has two main components: a hot,
fast and tenuous outflow along the jet axis and a cold, slow and dense flow
along the funnel wall defining the jet geometry (see also Meliani \& Keppens 2009).
For slow rotating BHs, the flow energetics is dominated by the
kinetic energy and the enthalpy of matter whereas for fast rotating BH the energetics
is essentially given by a Poynting flux. Jets
dominated by kinetic energy penetrate easily in the intra-cluster medium
(ICM), forming low density cavities elongated in the radial direction
(Guo \& Mathews 2011). This is not the case if the energetics of the jet is dominated
by highly relativistic particles. In this case, due to the low inertia,
the jet decelerates rapidly in the ICM, producing large cavities due to
the lateral expansion produced by the pressure of the relativistic particles
(Guo \& Mathews 2011).

The result of these simulations emphasize the importance of the presence of
relativistic particles in the jet, either to characterize the dynamics of
the flow or the interaction with the ICM. However, the acceleration mechanism
(or mechanisms) of these relativistic particles is not yet well understood
although magnetohydrodynamic shocks and Fermi-like mechanisms are often
invoked as possibilities. Here we examine a completely different
alternative, i.e., a purely gravitational acceleration process based on the
presence of an ergosphere around a Kerr BH. In our scenario, we assume
that an accreting rotating BH is present in the center of an active galaxy.
Matter penetrating the ergosphere can undergo the Penrose process (Penrose 1969) and, under
certain conditions, the emerging particles follow geodesics asymptotically parallel to the
rotation axis, acquiring very high energies. Although the efficiency of the Penrose process
is still a matter of debate, this important question will be not examined in the present paper.

The existence of unbound geodesics leaving the ergosphere along the $z$-axis and focusing at
infinity was already demonstrated by Gariel et al. (2010, hereafter GMMS10). In the present work,
it is shown that test particles, independently of their electric charge, following
these highly collimated geodesics, model a narrow energetic beam in precise regions of the ergosphere.
The present investigation is addressed to the study of some particular solutions describing those
geodesics as well as to the analysis of the
initial conditions required for their existence. The paper is organized as follows: in Section 2, the
constants of motion, the total particle energy $E$ and the Carter constant ${\cal Q}$ are
derived as functions of two real roots of the characteristic equation $R^{2}(r)=0$, where the function $R$
governs the timelike geodesics in the Kerr's metric; Section 3 is dedicated to the study of
the particular case in which a double root of the equation $R^2(r)=0$ exists. It is shown that high
particle energy values are possible only for two narrow ranges of values of the considered double root;
in Section 4 the two remaining
roots of the characteristic equation are examined as well as the
consequences for the allowed values of $E$ and for the asymptotes of the
geodetic motion; in Section 5 an analysis of these various solutions is performed,
permitting to restrict to one the different possibilities and, finally, in
Section 6 the main conclusions are given.

\section{The constants of motion}

Firstly, it should be emphasized that our model is highly idealized since
interactions with the ambient medium or with magnetic fields, which affect the
motion and the energy budget of particles, are not included in the present approach and will be
considered in a future paper.
This investigation will be focused on the
study of the motion of test particles following unbound geodesics along the
$z$-axis, under the assumption that the central supermassive black hole
(SMBH) rotates steadily.

Assuming an axisymmetric geometry, the generalized cylindrical or Weyl
coordinates ($\rho $, $z$, $\phi $), related to Boyer-Lindquist generalized
spherical coordinates ($r$, $\theta $, $\phi $) by
\begin{equation}
\rho =[(r-1)^{2}-A]^{1/2}\sin \theta ,\;\;z=(r-1)\cos \theta
\label{eq1}
\end{equation}
where
\begin{equation}
A=1-\left( \frac{a}{M}\right) ^{2}
\label{1b}
\end{equation}%
are the most suitable for describing the system. As already mentioned, the existence of special
unbound geodesics in this frame was recently demonstrated by GMMS10.
These geodesics stem from the ergosphere and when $z\rightarrow \infty$,
they are asymptotically parallel to the $z$ axis and positioned at distance
\begin{equation}
\rho =\rho _{1}\equiv \left( \rho _{e}^{2}+\frac{{\cal Q}}{E^{2}-1}\right)
^{1/2}  \label{1a}
\end{equation}
that depends on $\rho _{e}\equiv a/M$ and on two
constants of motion; the Carter constant ${\cal Q}$ and the energy $E$, while the
third constant of motion, the $z$ component of the angular momentum $L_{z}$
is necessarily null.

The function $R(r)$ (see, for instance, Chandrasekhar 1983) introduced in the expression of the
Kerr timelike geodesics (test particle mass $\sqrt{\delta _{1}}=1$) plays a
fundamental role in the analysis of the jet collimation when the engine at
the centre of the accretion disk is supposed to be a stationary rotating
SMBH. This function is a fourth order polynomial, i.e.,
\begin{equation}
R^{2}(r)=a_{4}r^{4}+a_{3}r^{3}+a_{2}r^{2}+a_{1}r+a_{0}
\label{1}
\end{equation}
where the coefficients, excepting $a_3$, depend on the constants of motion and on the
BH parameters (Chandrasekhar 1983) as
\begin{eqnarray}
a_{0}=-a^{2}{\cal Q},\;\;a_{1}=2(a^{2}E^{2}+{\cal Q})  \nonumber \\
a_{2}=a^{2}(E^{2}-1)-{\cal Q},\;\;a_{3}=2,\;\;a_{4}=E^{2}-1
\label{2}
\end{eqnarray}
Without loss of generality, we have put $M=1$ and $L_{z}=0$ when considering the special 2D-geodesics
given by Equation (\ref{1a}). Hence, the spin $a$ of the SMBH being fixed ($-1\leq a\leq 1$),
we have two independent parameters left, ${\cal Q}$ and $E$, or,
equivalently from Equation (\ref{1a}), the position $\rho _{1}$ of the asymptote
parallel to the $z$-axis and the energy $E$.

Let us consider the possible roots of the equation $R^{2}(r)=0$ of the
characteristics $\dot{r}=0$ of the autonomous system of geodesics equations
(Chandrasekhar 1983), i.e.,
\begin{equation}
a_{4}r^{4}+a_{3}r^{3}+a_{2}r^{2}+a_{1}r+a_{0}=0
\label{5}
\end{equation}
The polynomial given by Equation (\ref{5}) has four roots, labeled $r_{i}$ with
$i=1,2,3,4$, which can be {\it a priori} real (positive or negative)
or complex (contrary to the physical variable $r$, which is always real and defined in the interval
$\left[ 1+\sqrt{A},\infty \right[ $). The two equations $R^{2}(r_{1})=0$ and
$R^{2}(r_{2})=0$ are linear in ${\cal Q}$ and $(E^{2}-1)$. Thus, the solution of this linear
system permits to express the constants of motion as functions
of the roots $r_{1}$ and $r_{2}$, namely
\begin{equation}
{\cal Q}=\frac{2r_{1}r_{2}}{D}\left\{ a^{4}+a^{2}\left[
r_{1}(r_{1}-2)+r_{2}(r_{2}-2)\right]+r_{1}^{2}r_{2}^{2}\right\}
\label{6}
\end{equation}
and
\begin{equation}
(E^{2}-1)= -\frac{2}{D}\left\{a^{4}+a^{2}(r_{1}^{2}+r_{2}^{2})+r_{1}r_{2}
\left[r_{1}(r_{2}-2)-2r_{2}\right] \right\}
\label{7}
\end{equation}
with the denominator $D$ given by
\begin{equation}
D=a^{4}(2+r_{1}+r_{2})
+a^{2}\left[r_{1}^{3}+r_{1}^{2}r_{2}+r_{1}r_{2}(r_{2}-4)+r_{2}^{3}\right]+r_{1}r_{2}\left[ (r_{1}^{2}+r_{1}r_{2})(r_{2}-2)-2r_{2}^{2}\right]. \label{7aa}
\end{equation}

In the third possible equation, $R^{2}(r_{3})=0$, the parameters ${\cal Q}$
and $E^{2}-1$ can be replaced by Equations (\ref{6}) and (\ref{7}), leading to a
relation between $r_{3}$, $r_{1}$ and $r_{2}$, allowing in principle, to
determine the values of $r_{3}$ as a function of $r_{1}$ and $r_{2}$
only, since the spin parameter $a$ is fixed. The fourth possible equation, $R^{2}(r_{4})=0$,
does not represent any new result since the roots $r_{3}$ and $r_{4}$ are the
same.

In Equations (\ref{6}) and (\ref{7}), it is worth noting the symmetric role of $r_{1}$
and $r_{2}$, and that ${\cal Q}$ and $E^{2}-1$ have the same denominator
$D$. Thus, if and only if, it cancels, we have both $E\rightarrow \infty $ and
$\left\vert {\cal Q}\right\vert \rightarrow \infty$, whereas $\rho _{1}$, since it
depends only on their ratio (see Equation (\ref{1a})), tends towards a finite
value. From Equations (\ref{1a}), (\ref{6}) and (\ref{7}) we obtain for the position of the asymptotes
\begin{equation}
\left( \frac{\rho _{1}}{\rho _{e}}\right) ^{2}=\frac{(a^{2}+r_{1}^{2})(a^{2}-r_{1}r_{2})(a^{2}+r_{2}^{2})}{a^{2}
\left\{a^{4}+a^{2}(r_{1}^{2}+r_{2}^{2})+r_{1}r_{2}\left[ r_{1}(r_{2}-2)-2r_{2}
\right]\right\}}
\label{8}
\end{equation}
It should be emphasized that the roots $r_1$ and $r_2$, satisfying the condition $D=0$, for a
given value of the spin parameter $a$, permit to define regions in the ergosphere from which
unbound geodesics
parallel to the z-axis emerge, along which move particles with ``quasi"-infinite energies and that could
eventually explain the very energetic particles observed in cosmic rays. We will return to this
point later.

\section{Real roots $r_1 = r_2$}

In order to simplify the mathematical analysis and without loss in the physical insight,
we assume in this paper the particular case
in which a double real root $r_{1}=r_{2}=Y$ exists. Under this assumption Equation (\ref{1}) can be recast as
\begin{equation}
R^{2}(r)=a_{4}(r-Y)^{2}(r^{2}+Br+C)
\label{9}
\end{equation}
and Equations (\ref{6}) and (\ref{7}) simplify as
\begin{equation}
{\cal Q}=\frac{[a^{4}+2a^{2}(Y-2)Y+Y^{4}]Y^{2}}
{a^{4}(1+Y)+2a^{2}(Y-1)Y^{2}+(Y-3)Y^{4}}
\label{10}
\end{equation}
and
\begin{equation}
E^{2}-1= -\frac{a^{4}+2a^{2}Y^{2}+(Y-4)Y^{3}}
{a^{4}(1+Y)+2a^{2}(Y-1)Y^{2}+(Y-3)Y^{4}}
\label{11}
\end{equation}
When $E\rightarrow \infty$, of course we have also $\left\vert{\cal Q}\right\vert \rightarrow
\infty$ but, as already mentioned, their ratio tends towards a finite
value in order that the coordinate $\rho _{1}$ remains finite, i.e.,
\begin{equation}
\left( \frac{\rho _{1}}{\rho _{e}}\right)^{2}=
\frac{(a^{2}-Y^{2})(a^{2}+Y^{2})^{2}}{a^{2}[a^{4}+2a^{2}Y^{2}+(Y-4)Y^{3}]}
\label{12}
\end{equation}

In order to perform some numerical estimates, we will assume, unless otherwise stated, a "moderate" rotation for
the SMBH, fixing $a=M/2$ (a value
adopted already by GMMS10 in their investigation). Hence, the functions
$E^{2}-1 \equiv F(Y)$ and $(\rho _{1}/\rho _{e})^{2} \equiv G(Y)$ can be plotted as shown in Figures 1 and 2.

%%%%%%%%%%%%%%%%%%%%%%%%%%%%%%%%%%%% figure 1%%%%%%%%%%%%%%%%%%%%%%%%%%%%%%%%%%%%%%%%%%%%%%%%%
\begin{figure}
\epsscale{.80}
\plotone{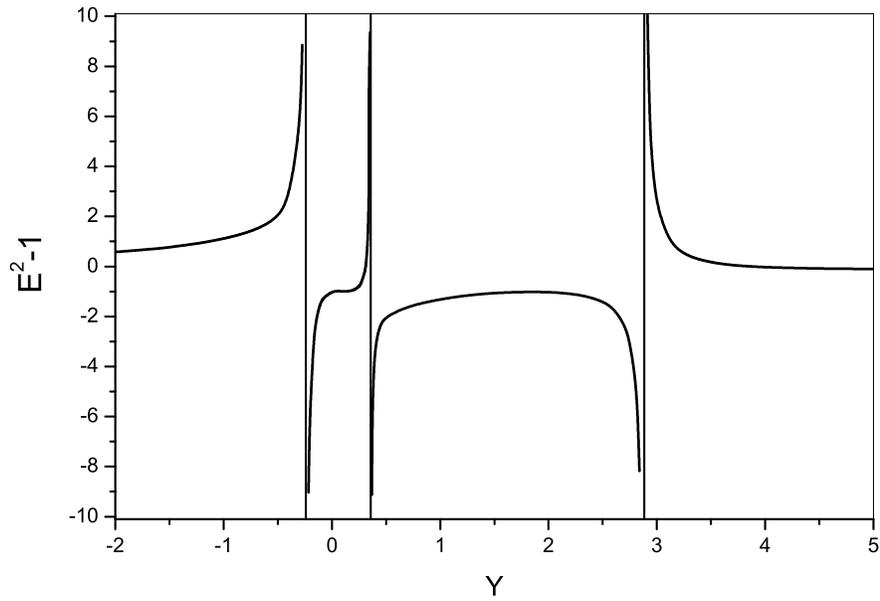}
\caption{Plot of $E^{2}-1 \equiv F(Y)$, where $E$ is the energy of the test-particle as a
function of the double root $Y$, computed from Equation (13) for a BH of mass $M=1$ and spin parameter
${a}/{M}=0.5$. The values of $Y$ for which $E^{2}-1$ is positive, as expected for
unbound geodesics, are clearly seen as well as the three
values of $Y$ for which $E^{2}-1$ tends (positively) to infinity.}
\end{figure}
%%%%%%%%%%%%%%%%%%%%%%%%%%%%%%%%%%%%%%%%%%%%%%%%%%%%%%%%%%%%%%%%%%%%%%%%%%%%%%%%%%%%%%%%%%%%%%%%%%%%%%%%%%%%%%%%

Since $F(Y)$ and $G(Y)$ must be simultaneously positive, the only possible
solutions correspond to the two intervals
\begin{equation}
Y\in\left[ -0.5,Y_{0a}\right]
\label{13}
\end{equation}
and
\begin{equation}
Y\in\left[ Y_{0b},3.8697\right]
\label{13a}
\end{equation}
with $Y_{0a}$\ and $Y_{0b}$\ being asymptotes of $F(Y)$ for which
$E\rightarrow \infty$. These asymptotes can be evaluated numerically since they correspond to
the roots of the equation $D=0$. For the assumed BH parameters, it results
$Y_{0a}\simeq -0.2418$\ and $Y_{0b}\simeq 2.8832$.

Hence, there are only two possible values of $\rho _{1}$ for which $E\rightarrow \infty $, corresponding
to the two intervals defined above. For these limits, $Y=Y_{0a}-\varepsilon $ and $Y=Y_{0b}+\varepsilon $, when
$\varepsilon\rightarrow 0$, we obtain respectively for the coordinate $\rho_1$
\begin{equation}
\frac{\rho _{1}}{\rho _{e}}\simeq 0.6932 \;\;\mbox{and}\;\;\frac{\rho _{1}%
}{\rho _{e}}\simeq 10.2411.
\label{14}
\end{equation}

For the upper bound of the interval (\ref{13a}), i.e., for $Y=3.8697$ where
$\rho_{1}\rightarrow \infty$, we have $E^{2}-1=0$  and, for the lower bound of the interval
(\ref{13}), i.e., for $Y=-0.5$ where $E^{2}-1=2$, we
have $\rho_{1}=0$.

%%%%%%%%%%%%%%%%%%%%%%%%%%%%%%%%%%%% figure 2%%%%%%%%%%%%%%%%%%%%%%%%%%%%%%%%%%%%%%%%%%%%%%%%%
\begin{figure}
\includegraphics[angle=0,scale=.50]{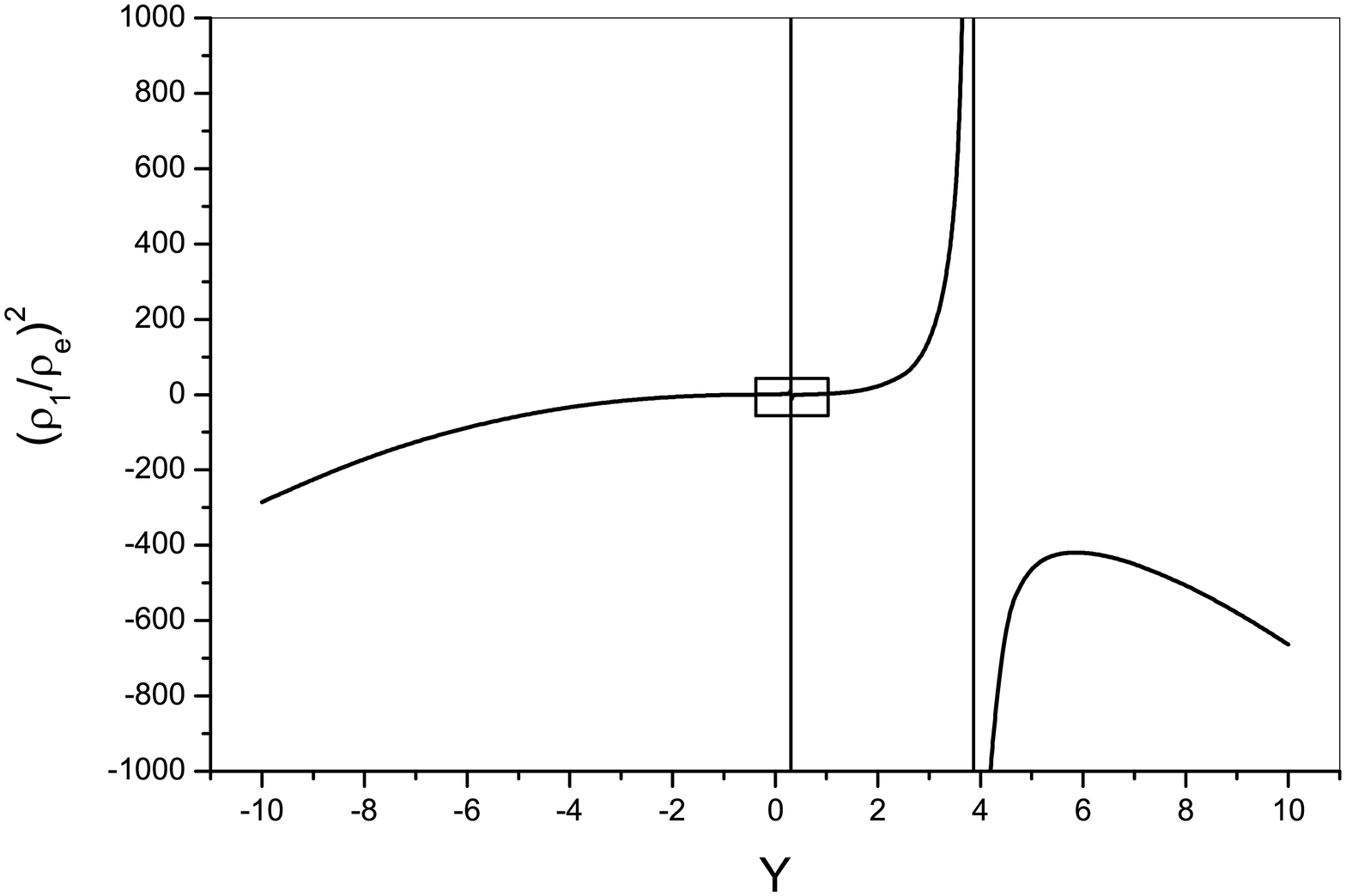}
\includegraphics[angle=0,scale=.50]{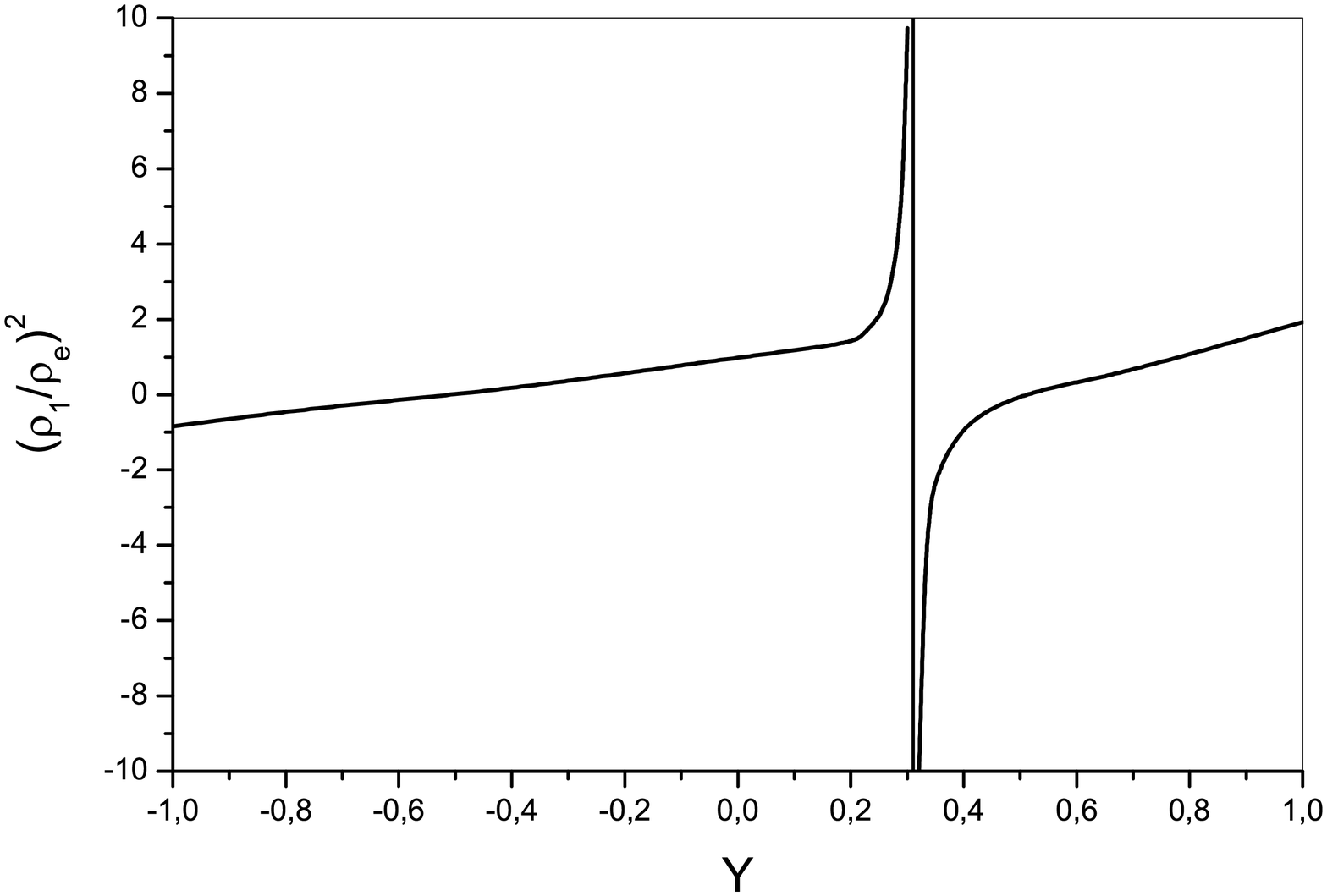}
\caption{{\bf Upper panel}: Plot of the ratio $(\rho_1/\rho_e)^2$ as function of the double root $Y$, evaluated
from Equation (\ref{12}), for a BH of mass $M=1$ and spin parameter ${a}/{M}=0.5$.
The intervals of $Y$ in which the ratio $G(Y)$ is positive can be seen. {\bf Lower panel}: details of the
variation of the ratio $G(Y)$ corresponding to the rectangle shown in the upper panel.}
\end{figure}
%%%%%%%%%%%%%%%%%%%%%%%%%%%%%%%%%%%%%%%%%%%%%%%%%%%%%%%%%%%%%%%%%%%%%%%%%%%%%%%%%%%%%%%%%%%%%%%%%%%%%%%%%%%%%%%%

When $r_1 \not= r_2$, the problem is quite difficult and a more careful study is necessary.
This is presently underway and will be reported in a future paper.
Nevertheless, the following aspects may be anticipated. As we have seen, in the case of a double
root, geodesics followed by particles of sufficient high energy
($E>3$), remain clustered around values of $\rho_1$ corresponding to
$E\rightarrow\infty$. Moreover, in the general case the condition $D(r_1,r_2)=0$ defines the
regions from which high energy geodesics emerge. The later condition is given by
a third order polynomial in  $r_2$, which has a unique real root that can be explicited as a function
of $r_1$. This can be inserted into Equation (\ref{8}) defining the position of the asymptote $\rho_1$ as a function
only of $r_1$. That leads again to very restricted ranges of possible values of $\rho_1$.

For the sake of completeness and in order to investigate also the possible influence of
the spin parameter in our analysis, we have also considered the case of a extreme Kerr black hole
($a/M=1$). The results are qualitatively the same. The only physical solution leading to $E\rightarrow\infty$
corresponds to the double root $Y$=-0.4142 and to an asymptote whose position is $\rho_1\simeq 0.8284$.

\section{Roots $r_3$ and $r_4$}

Identifying Equation (\ref{9}) with Equation (\ref{1}), rewritten in terms of the parameters
$E^2-1=F(Y)$ and $(\rho_1/\rho_e)^2=G(Y)$, without the explicit form of these
functions of $Y$ (see Equations (\ref{11}) and (\ref{12})), yield the four
relations,
\begin{eqnarray}
B-2Y=\frac{2}{F}, \;\; G+\frac{1}{F}=2Y(BY-2C),  \nonumber \\
1-G=16CY^2, \;\; 2-G=4(C-2YB+Y^2),
\label{16}
\end{eqnarray}
which are linear in $1/F$, $G$, $B$ and $C$.
After eliminating $1/F$ and $G$, we obtain from the equations above
\begin{eqnarray}
B=-\frac{2(4Y^2-1)[Y(4Y-1)+Y-1]}{(4Y^2-1)^2-16Y^2(4Y-1)},  \label{17} \\
C=\frac{(4Y^2-1)^2+16Y(Y-1)}{4[(4Y^2-1)^2-16Y^2(4Y-1)]}.  \label{18}
\end{eqnarray}
Hence Equation (\ref{9}) can be recast as
\begin{eqnarray}
R^2=a_4(r-Y)^2(r^2-Sr+P) \\ \nonumber
=a_4(r-Y)^2(r-r_3)(r-r_4),
\label{19}
\end{eqnarray}
where $r_3$ and $r_4$ are the remaining roots, in general distinct, and we have introduced
\begin{equation}
S\equiv r_3+r_4=-B, \;\; P\equiv r_3r_4=C,
\label{20}
\end{equation}
or
\begin{eqnarray}
r_3=-\frac{1}{2}\left[B+\left(B^{2}-4C\right)^{{1}/{2}}\right],  \label{21}
\\
r_4=-\frac{1}{2}\left[B-\left(B^{2}-4C\right)^{{1}/{2}}\right],  \label{21a}
\end{eqnarray}
where $B(Y)$ and $C(Y)$ are given respectively by Equations (\ref{17}) and (\ref{18}).

The curves $r_{3}(Y)$ and $r_{4}(Y)$ are real only for some well defined ranges of $Y$. In particular,
in the interval (\ref{13}) defining $Y$, $r_{3}$ and $r_{4}$ are not real. In order to have
the expression $r^{2}+Br+C$ in Equation (\ref{9}) real, where $B$ and $C$ are
real, $r_{3}$ and $r_{4}$ have to be complex conjugated, i.e.,
$r_{3}=z=B_{1}+iC_{1}$ and $r_{4}=\overline{z}$. Hence, the sign of the
expression $r^{2}+Br+C=(r+B_{1})^{2}+C_{1}^{2}$ is always positive, and
$P=C=B_{1}^{2}+C_{1}^{2}\geq 0$ and $S=-B=-2B_{1}\leq 0$.

In the interval (\ref{13a}), the two roots $r_{3}$ and $r_{4}$ are
real, $P=C$ is negative (which means two roots of opposite signs) and $B=-T$
is positive. The most precise value we have numerically obtained
for the left limit (where, in principle, $E\rightarrow \infty $) of the
range (\ref{13a}) allows us to reach the value $E\simeq 1.1\times
10^{32}$. Then, the corresponding real values of the roots are respectively
$r_{3}=-5.8988$ and $r_{4}=0.1324$.

Also, it is worth observing that $E$ is steeply decreasing either for a weak
variation $\varepsilon $ of $Y$ from $Y_{0b}$ \ ($\varepsilon >0$) or from
$Y_{0a}$ ($\varepsilon <0$), while $\rho _{1}$ is weakly increasing for this
same small interval of $Y$. For example, when $Y$ goes from $Y_{0b}$ to $%
2.922$, the energy $E$ is steeply decreasing from $10^{30}$ to $3$, while
the position of the asymptote $\rho _{1}/\rho _{e}$ increases by a small
amount from $10.24$ to $10.68$, which means a large concentration of the most
energetic part (the "spine") of the beam immediately near at the right hand
side of $\rho _{1}/\rho _{e}=10.24$. At its left hand side, there is no
beam produced.

Likewise, for the interval (\ref{13}), the energy $E$ is very steeply
decreasing from ``infinity" to $6$. Within our numerical precision,
when $Y\rightarrow Y_{0a}$, the highest value obtained for the energy was
$E_{max}\simeq 2\times 10^{30}$. The corresponding asymptote
$\rho_{1}/\rho _{e}$ inside the ergosphere decreases very slightly from $0.6932$
to $0.6764$. Here the jet is still more concentrated just at the left of coordinate value
$0.6932$, while beyond its right side, there is no beam at all.

As a result, the present model predicts a radial structure of the ``jet", with
a well defined profile for the energy (or velocity) distribution of the particles.

\section{Unbound geodesics for high energy particles}

Now let us consider the set of possible energetic geodesics framing a jet and
their corresponding asymptotes, satisfying the conditions  considered in the previous section.
In this case, the choice of admissible initial conditions (IC) depends strongly
on their positions relative to the characteristics of the system of geodesic
equations (see Equations (2) and (3) in GMMS10). Indeed, each characteristic separates
the plane into two regions and a given geodesic cannot cross the borders
defined by those curves. In Boyer-Lindquist
coordinates, the characteristics are defined by the equations
\begin{equation}
{\dot r}=0, \;\; {\dot \theta}=0,
\label{39}
\end{equation}
each of which is equivalent to the
equations,
\begin{equation}
R(r)=0, \;\; T(\theta)=0,  \label{40}
\end{equation}
respectively, where $R(r)$ is given by Equation (\ref{1}) and $T(\theta)$ by
\begin{equation}
T(\theta)=-(b_4\mu^4+b_2\mu^2+b_0)^{1/2}, \label{x}
\end{equation}
with $\mu=\cos\theta$ and
\begin{equation}
b_0=\frac{\mathcal Q}{M^2}, \;\; b_2=a_2, \;\; b_4=-\left(\frac{a}{M}\right)^2(E^2-1). \label{xx}
\end{equation}
The solutions of Equation (\ref{40}), when they exist, are the roots $r_i$
and the roots $\theta_i$, which define respectively circles and straight lines from the
origin.

In Weyl coordinates $\rho $ and $z$, each characteristic Equation (25)
is equivalent to the equations (see Equations (17) and (18) in GMMS10)
\begin{equation}
{\dot \rho }=\frac{R(\alpha^2-A)z}{\alpha \rho \Delta}, \;\; {\dot z}=
-\frac{R\alpha }{\Delta},  \label{41}
\end{equation}
and
\begin{equation}
{\dot \rho }=\frac{T\alpha ^{3}\rho }{(\alpha ^{2}-A)\Delta}, \;\; {\dot z}=%
\frac{T\alpha z}{\Delta},  \label{42}
\end{equation}
respectively, where
\begin{equation}
\Delta=(\alpha+1)^{2}\alpha ^{2}+\left(\frac{a}{M}\right)^{2}z^{2}.
\label{42a}
\end{equation}
Each set of Equations (\ref{41}) and (\ref{42}) leads to
\begin{equation}
\frac{dz}{d\rho }=-\frac{\alpha^2\rho}{(\alpha^2-A)z},  \label{43}
\end{equation}
and
\begin{equation}
\frac{dz}{d\rho }=\frac{(\alpha^2-A)z}{\alpha^2\rho },  \label{44}
\end{equation}
respectively, defining the two families of characteristics for the geodesics
of type Equation (\ref{39}) in which we are interested, namely, ellipses
(corresponding to ${\dot r}=0$) and hyperbolae (corresponding to ${\dot
\theta }=0$). Let us note that the product of the derivatives of Equations (\ref{43})
and (\ref{44}), of these two characteristics is $-1$, indicating that they
are orthogonal.

In the one hand ellipses exist when there are solutions $r=r_i=$ constant
of Equation (\ref{43}) for any $\theta$, with $r_i\geq 1+\sqrt{A}$ or
equivalently $\alpha =\alpha_i=$ constant ( because $\alpha =r-1$) with $%
\alpha_i\geq \sqrt{A}$. Then Equation (\ref{43}) can be integrated yielding
\begin{equation}
\left(\frac{z}{\alpha_i}\right)^2+\frac{\rho^2}{\alpha_i^2-A}=K_1,
\label{45}
\end{equation}
where $K_1$ is an integration constant. The comparison
of Equation (\ref{45}), valid for any $\theta$, with Equation (\ref{eq1}), implies $K_1=1.$

On the other hand, hyperbolae exist when there are solutions of $\cos\theta\equiv\mu =\mu_i=$
constant of Equation (\ref{44}) for any $r$, with $\mu_i^2\leq 1$. These are
solutions of the equation $T=0$, when $L_z=0$, with
\begin{equation}
T^{2}=\left(\frac{a}{M}\right)^{2}(E^{2}-1)\alpha^4 \left[1-\left(\frac{z}
{\alpha }\right)^2\right]\Lambda,
\label{46}
\end{equation}
where we have defined $\Lambda=\left\{{\cal Q}/\left[a^{2}(E^{2}-1)\right]+\left(z/\alpha\right)^2\right\}$.
There are two possible cases, namely $\mu^2_i=1$, then $T=0$ for any
${\cal Q}$, or $\mu_{i}^{2}=-{\cal Q}/[a^2(E^2-1)] =1-(\rho_1/\rho_e)^{2}\leq 1$,
being positively defined only if ${\cal Q\leq }0$, or equivalently, if
$\rho_1\leq \rho_e$. Then, for $\mu^2_i=1$, we have $z=r-1=\alpha $ and $\rho
=0$ for any $r$ and Equation (\ref{44}) reduces to $\rightarrow \infty $\ , and the
characteristics being along the semi-axis $z\geq \sqrt{A}$. For
$\mu_i^2=-{\cal Q}/[a^{2}(E^2-1)]$ we have from Equation (\ref{44})
\begin{equation}
\frac{dz}{d\rho }=\frac{z^2-A\mu_i^2}{z\rho},  \label{47}
\end{equation}
which can be integrated leading to
\begin{equation}
\rho=K_2\left[\left(\frac{z}{\mu_i}\right)^2-A\right]^{1/2},  \label{48}
\end{equation}
where $K_2$ is an integration constant. Comparing Equation (\ref{48}), valid for any $r$, or equivalently for
any $\alpha$, with Equation (\ref{eq1}) yields $K_2+\mu_i^2=1$.

Relation (\ref{48}) represents a family of hyperbolae parametrized by
\begin{equation}
\frac{\rho_1}{\rho_e}=\left(1-\mu_i^2\right)^{1/2}
\end{equation}
yielding
\begin{equation}
\frac{1}{A}\left[1-\left(\frac{\rho_1}{\rho_e}\right)^2\right]^{-1}z^2-
\frac{1}{A}\left(\frac{\rho_1}{\rho_e}\right)^{-2}\rho^2=1
\label{49}
\end{equation}

If the IC of a geodesic lies inside an ellipse of the
type  defined by Equation (\ref{45}), such a geodesic cannot be unbounded and hence
cannot go to infinity. So, the allowed IC has to satisfy a triple
condition, i.e., i) be inside the ergosphere in order to be possibly issued
from a Penrose process; ii) be outside the largest elliptic
characteristic, corresponding to the larger value of the roots
$r_i$ and iii) be above the higher hyperbolic characteristic, which
corresponds to the highest value of the roots $\left\vert \mu_i\right\vert
\leq 1$. These conditions restrict the permitted domain of IC.

An ellipse, solution of Equation (\ref{45}), when it exists (i.e. when
$r_i\in [1+\sqrt{A},\infty [$), can intersect the ergosphere only
if its semi-minor axis $b_{i}=(\alpha_i^2-A)^{1/2}$ is smaller
than $\rho_e=a/M$, i.e. if $r_i<2$.

As an example, let us consider again the special case of a double root $Y$ studied previously.

a) The first admissible range is $Y\in \lbrack Y_{0b},3.869]$
(see Figure 2). These roots, belonging to the domain of
physical definition $r\in [ 1+\sqrt{A}=1.86,\infty [$,
correspond to the existence of elliptic characteristics. The smallest
ellipse has as semi-minor axis
$b_{i}=[(Y-1)^{2}-A]^{1/2}=[(1.88)^{2}-0.75]^{1/2}=(2.78844)^{1/2}$ along
$\rho $, and as semi-major axis $a_{i}=\sqrt{\alpha _{i}^{2}}=Y-1=1.88$ along
$z$, obtained for the smallest value $Y_{0b}\simeq 2.88$, corresponding to
$\rho _{1}/\rho _{e}\simeq 10.24$. This ellipse contains the ergosphere, the
limits of which being $z_{\max }=\sqrt{A}=0.866$ and $\rho _{\max }=1/2$.
Hence, it is always impossible to have IC simultaneously inside the
ergosphere and outside any ellipse. Thus, in this case, it is impossible
to have unbound geodesics starting from the ergosphere.

b) For the second admissible range, we found that $Y\in \lbrack -0.5,Y_{0a}]$
(see Figure 2). These roots do not belong to the domain of
definition of the physical variable $r$, which means that there is
{\bf no} corresponding elliptic characteristic. The only remaining possible
limitation depends on the position of the hyperbolic characteristics Equation
(\ref{49}). The hyperbola intersects the $z$ -axis at the point with coordinates
$\rho =0$ and $z_{0}=\{A[1-(\rho _{1}\rho _{e})^{2}]\}^{1/2}$ and tends
asymptotically towards a straight line of equation $\rho \simeq z\tan
\theta _{1}$, with $\sin \theta _{1}=\rho _{1}/\rho _{e}$. The domain of
possible IC is located between the $z$-axis, the ergosphere and
the hyperbola. For example, for $Y=-0.242$, $\rho _{1}/\rho
_{e}=0.693$, $\theta _{1}=21^{o}$, and $z_{0}=0.624(<\sqrt{A}=0.866)$.

{\bf As we have seen, geodesics characterized by $E\rightarrow \infty $ may exist. Indeed,
trajectories defined by the differential Equation (21) in GMMS10, are
well behaved when $E\rightarrow\infty$ (but with $\rho_1$ finite), 
as can be easily verified by factorizing $E$ in the
numerator and the denominator of the second member, leaving a finite
quantity when $E\rightarrow\infty$. The existence of such trajectories has been verified
by the numerical solution of Equation (21) in GMMS10 for high energy values
and the corresponding values of $\rho_1$ as described in Section 3. 
Of course, the value of the energy will be always finite and fixed by the particular
Penrose process which takes place at the origin of these particular geodesics inside
the ergosphere. The main point resulting from the present investigation is that if very
energetic particles are produced as a consequence of such a process, there are unbound
geodesics that will be followed by those particles, which will permit their ejection 
in a collimated way along the spin axis of the black hole.} 

In Figure 3 is plotted the geodesic which tends asymptotically
towards the corresponding value $\rho =\rho _{1}=0.347$, for which the test
particle has a very high (theoretically ``infinite",but for present
calculations, we took value $E=10^{6}$ and $\rho _{1}/\rho
_{e}=0.693$). This plot corresponds to the IC $\rho _{i}=2.8\times
10^{-6} $ and $z_{i}=0.8521$, which correspond to a point just inside the
ergosphere at its top near the $z$-axis, i.e., near the event horizon. For
the other limit ($Y=-0.5$), $\rho _{1}=0$, $E=\sqrt{3}$,
$\theta_{1}=0$ and $z_{0}=\sqrt{A}=0.866$.

%%%%%%%%%%%%%%%%%%%%%%%%%%%%%%%%%%%%%%%%%%%%% figure 3 %%%%%%%%%%%%%%%%%%%%%%%%%%%%%%%%%%%%%%%%%%%%%%%%
\begin{figure}
\includegraphics[angle=0,scale=.50]{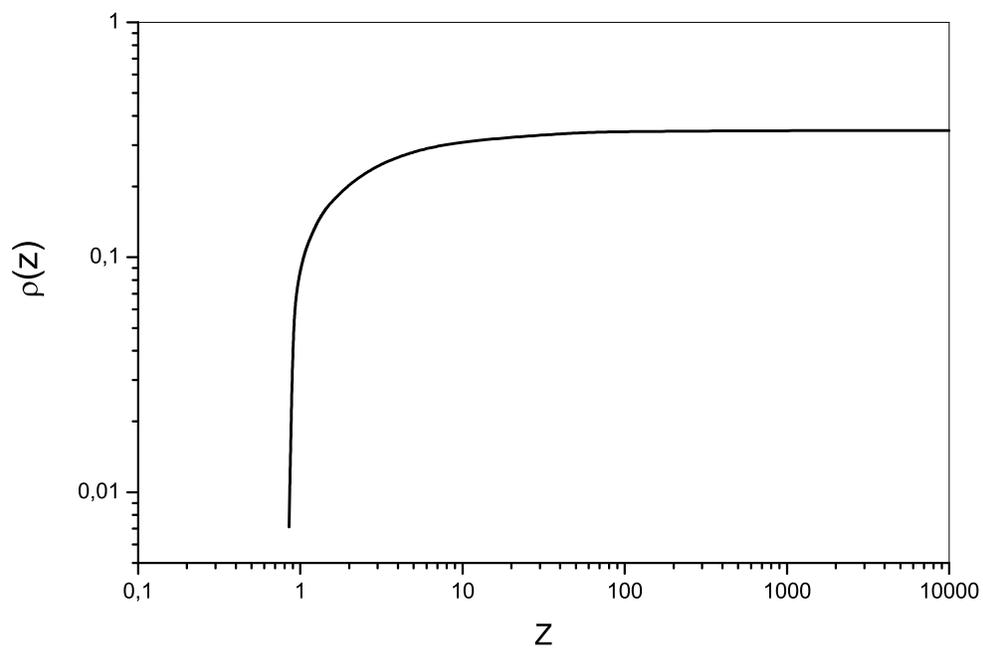}
\caption{Plot of the geodesic $\rho(z)$ for a Kerr BH characterized by $M$ = 1 and
a spin parameter $a/M$ = 0.5. The constants of motion and initial conditions are those given in the text.}
\end{figure}
%%%%%%%%%%%%%%%%%%%%%%%%%%%%%%%%%%%%%%%%%%%%%%%%%%%%%%%%%%%%%%%%%%%%%%%%%%%%%%%%%%%%%%%%%%%%%%%%%%%%%%%%%%%%%%%%%%%

In order to illustrate the considered geometry, we have plotted in Figure 4 the relevant curves
and surfaces like the ergosphere, the critical hyperbola, a possible unbound geodesic and its
corresponding asymptote. Curves in black and with subscript $a$ correspond to the case $a/M=0.5$ considered above, while
curves in red and with subscript $b$ correspond to the extreme case $a/M=1$. The geodesic for the case $a/M=1$ was 
calculated for the following values of the constants: $E=10^6$ and $\rho_1=0.8284$ (see end of Section 3) and for 
the following IC, $\rho_i=0.1$ and $z_i=0.1675$. Notice that in the
extreme case, the critical hyperbola, which delimitates with the ergosphere and the z-axis the region
where unbound geodesics emerge, degenerates into a straight line starting from the origin and having
a slope $\Delta z/\Delta\rho\simeq 0.676$. Moreover, the high energy beam of geodesics has a width,
measured by the distance between the z-axis and the asymptote, which is about a factor of two
larger than that obtained for the case $a/M=0.5$.

%%%%%%%%%%%%%%%%%%%%%%%%%%%%%%%%%%%%%%%%%%%%% figure 4 %%%%%%%%%%%%%%%%%%%%%%%%%%%%%%%%%%%%%%%%%%%%%%%%
\begin{figure}
\includegraphics[angle=0,scale=.50]{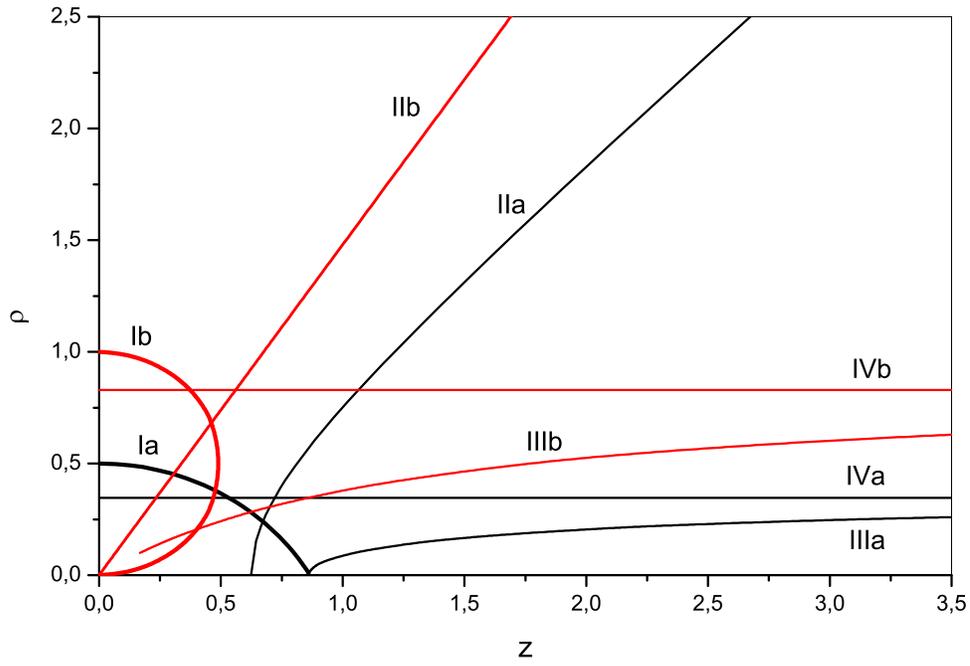}
\caption{The ergosphere (labels Ia and Ib), the critical hyperbola (labels IIa and IIb), an example of a unbound
geodesics (labels IIIa and IIIb) and the respective asymptotes (labels IVa and IVb) are shown for black holes
having a spin parameter $a/M$ = 0.5 (black curves) or $a/M$ = 1 (red curves).}
\end{figure}
%%%%%%%%%%%%%%%%%%%%%%%%%%%%%%%%%%%%%%%%%%%%%%%%%%%%%%%%%%%%%%%%%%%%%%%%%%%%%%%%%%%%%%%%%%%%%%%%%%%%%%%%%%%%%%%%%%%

{\bf Let us return to the case $a/M=0.5$.} By modifying a little bit the preceding IC, we find other 
geodesics similar to the latter ones but diverging or, in other words, deviating slightly from its
asymptote $\rho _{1}$. For instance, taking as IC $\rho _{i}=2.8\times
10^{-6}$ and $z_{i}=0.852086175$ we obtain a geodesic that diverges almost
rectilinearly from the turning point at $\rho \simeq 0.3466$ and $z\simeq
50\times 10^{3}$ attaining at $z=10^{10}$ (or 1 Mpc) the abscissa $\rho
\simeq 350$ (or $3\times 10^{-2}$ pc), which corresponds to a slight slope $%
\Delta \rho /\Delta z=3.5\times 10^{-8}$. This geodesic conserves its highly
collimated character. If this geodesic is followed by an electron, its
kinetic energy converges to the total energy $E$, attaining at $z=10^{10}$
(or 1 Mpc from the BH) a Lorentz factor $\Gamma =5\times 10^{4}$, which corresponds to
an energy 25 GeV. This same electron at $z=1.4\times 10^{9}$ (or 140 kpc),
where $\rho \simeq 50$ (or $5\times 10^{-3}$ pc), attains a Lorentz factor $%
\Gamma \simeq 1.9\times 10^{4}$ or an energy of the order $9.4$ GeV.

The local Lorentz factor along a given geodesic is defined by its usual expression
$\Gamma = (1-v^2/c^2)^{-1/2}$, where $v^2 = v^2_z + v^2_{\rho}$, $v_z = \dot z/\dot t$ and
$v_{\rho} = \dot\rho/\dot t$. The overdot means the derivative with respect to
the proper time $\tau$. Gravitational effects are included self-consistently in the expressions
defining the Kerr geodesics. In this way, $\Gamma$ can be expressed as a function of the coordinates
$\rho$ and $z$. For the case previously considered (geodesic shown in Figure 3), the variation of the
Lorentz factor along the geodesic can be expressed as a function of the
coordinate $z$ alone. The result is shown in Figure 5.
Notice that in the beginning the
energy is essentially under the form of potential energy ($\Gamma \simeq 1$), while far from
the ergosphere, the energy is essentially kinetic.

%%%%%%%%%%%%%%%%%%%%%%%%%%%%%%%%%%%%%%%%%%%%% figure 5 %%%%%%%%%%%%%%%%%%%%%%%%%%%%%%%%%%%%%%%%%%%%%%%%
\begin{figure}
\includegraphics[angle=0,scale=.50]{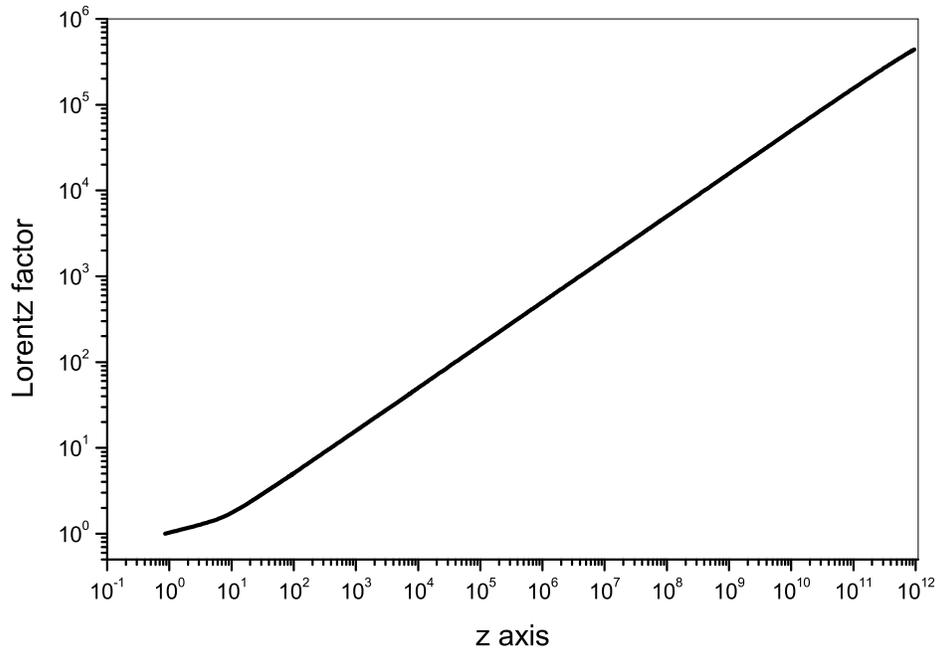}
\caption{Lorentz factor as a function of the coordinate $z$ for the geodesic shown in Figure 3.}
\end{figure}
%%%%%%%%%%%%%%%%%%%%%%%%%%%%%%%%%%%%%%%%%%%%%%%%%%%%%%%%%%%%%%%%%%%%%%%%%%%%%%%%%%%%%%%%%%%%%%%%%%%%%%%%%%%%%%%%%%%

\section{Conclusions}

Taking the roots of the characteristic equation for unbound 2D-geodesics
with $L_{z}=0$ as free parameters, we have shown that the two remaining
constants of motion, $E$ and ${\cal Q}$, of a test particle following
geodesics that asymptotically tend to a parallel line to the $z$-axis, can be
expressed as a function of the aforementioned parameters. In the particular case of a
double root and assuming a spin parameter to be
equal to $a=M/2$, restricted domains of asymptotes corresponding to high
particle energies are found. This means that the Kerr metric can generate
powerful collimated jets of high energy particles in some well defined regions inside the
ergosphere. Indeed, for this special case, only two possible ranges of $\rho
_{1}$ are possible, namely $\rho _{1}\in \left[ 0.3382,0.3466\right] $ and $%
\rho _{1}\in \left[ 5.12,5.34\right] $ for $E\in \left[ \sqrt{6},\infty %
\right[ $ and $E\in \left] \infty ,\sqrt{3}\right] $ respectively (see
Figure 2).

It is worth mentioning that the present results are a strict consequence of the
structure of the Kerr metric. The main approximation is the assumption
of the existence of a double real root $Y$ for the characteristic equation.
Although we have considered in some more detail the case of a moderately
spinning black hole ($a/M=0.5$), the results are qualitatively the same for the case of
a extreme Kerr black hole. We hope that by relaxing the double root hypothesis, maybe
it would be possible to obtain thicker beams of energetic particles.
An investigation is currently in progress and preliminary results are encouraging.

Two positions of the asymptote corresponding to an
"infinite" energy in our model are $\rho _{1}\simeq 5M$ and $\rho _{1}\simeq 0.3466M$. The latter is
the only case compatible with the limitations imposed by the
characteristics of the system of geodesic equations, according to our discussion in Section 5.
The consequence of these mathematical constraints is that the
resulting jet is very narrow. However, there is some observational
evidence for the presence of radial flows in some jets (Giroletti et al. 2004) and, as already
mentioned, they may have a two-component flow, i.e., a relativistic powerful
inner jet and a slower, less powerful outer flow (De Villiers et al. 2005; Xie et al. 2012). The
present model could be related to the inner flow which carries most of the
power, being constituted mainly by relativistic particles.

Our results can also easily be extended to particles other than electrons,
for example protons or heavy nuclei. This does not modify the "geometry"
that we obtained, i.e., the positions $\rho_1$ of the jets, but their energy
only. For a proton ($c^2\sqrt{\delta_1}\simeq$1 GeV) the maximal energy we
can here numerically calculate (which is theoretically as large as wanted)
is about $E\simeq 5.6\times 10^{25}$ eV $=5.6\times 10^7$ EeV, which largely
includes the highest energies of the current
observed UHECR (Dermer et al. 2009; Hoover et al. 2010).
Thus, our model could be relevant to explain not only a collimated flux of
relativistic particles but also the production of these very energetic particles
that could be related to UHECR. A detailed analysis of the Penrose
process and of its efficiency is beyond the aim of this paper. Nevertheless, we would expect
that the rate of emerging particles would be proportional to the accretion rate and to the ratio between
the volume of the region in the ergosphere where unbound geodesics exist and the total volume
of the ergosphere.
A simple example of a Penrose event could be the ionization of the inner shells of an iron atom
inside the ergosphere, with the nucleus being captured by the BH and the electron being ejected with
a high energy. The accreted gas near the ergosphere has temperatures around $10^6$ K (Montesinos \& de Freitas Pacheco 2011), high
enough to ionize the K-L shells of iron, providing a theoretical basis for such a possibility.

Recent results of the Pierre Auger Observatory (Roulet 2009) suggest a
correlation between UHECR above 57 EeV and nearby ($<70$ Mpc) AGNs.
Since in the present model the collimation of the relativistic particles
occurs along the spin axis, it would be interesting to investigate if the
associated AGNs are of type I or II, since in the context of the
``unified model" these classes differ only  by the
inclination of the jet axis with respect to the line-of-sight.

Let us recall briefly that since our model does not
require magnetic fields, it can be applied to neutral particles as, for
instance, neutrinos. If they have a mass of $\sqrt{\delta_1}=0.33$ eV
(Steidl 2009), according to our previous estimate, they could attain energies of the order of
$E\simeq 2\times 10^{-2} $ EeV.

Finally, different authors have recently discussed the possibility to produce high energy
particles by collisions near a rotating BH (Grib \& Pavlov 2011), raising the interest of
having further investigations on the Penrose process (see also Ba\~{n}ados at al 2011). The present
investigation can be seen as an additional contribution to this debate.

\end{document}